\documentclass{article}[12pt] 

\usepackage[english]{babel} 
\usepackage{amsmath}
\usepackage{amssymb}
\usepackage{txfonts}
\usepackage{mathdots}
\usepackage[classicReIm]{kpfonts}
\usepackage{graphicx}

\begin{document}

\begin{center}
\textbf{GRAVITATIONAL-SCALAR INSTABILITY OF A COSMOLOGICAL MODEL BASED ON A\\[12pt] 
TWO-COMPONENT SYSTEM OF DEGENERATE SCALARLY CHARGED FERMIONS\\[12pt] 
WITH ASYMMETRIC HIGGS INTERACTION. I. EQUATIONS FOR PERTURBATIONS.}\\[12pt]

Yu.~G.~Ignat'ev\\[12pt]

A mathematical model is formulated for the evolution of plane perturbations in a cosmological two-component statistical system of completely degenerate scalarly charged fermions with an asymmetric scalar Higgs interaction. A complete closed system of differential equations describing the unperturbed state of a homogeneous and isotropic system and a system of self-consistent evolution equations of small perturbations are constructed.

\end{center}

\section*{Introduction}

In recent years, in connection with the simultaneous direct experimental detection in 2016 of gravitational waves and Black Holes [1], [2] and their subsequent study, in particular, earlier indirect observations on the orbits of stars of the supermassive Black Hole at the center of our Galaxy were confirmed with a mass of the order of magnitude (see, for example, [3], [4]), as well as the existence of supermassive Black Holes at the centers of galaxies and quasars with masses in the range $10^{9} -10^{11} M_{\mathrm{\odot }} $, such as, for example, the supermassive Black Hole SDSS J140821.67+025733.2 at the center of a quasar SDSS J140821, which has a mass of $1.96\cdot 10^{11} M_{\mathrm{\odot }} $.

It is assumed that supermassive black holes with the mass of $\mathrm{\sim }10^{9} M_{\mathrm{\odot }} $, are central objects of luminous quasars observed at redshifts \textit{z} $\mathrm{>}$ 6, but their astrophysical origin remains not fully understood.  There are currently discovered over 700 quasars and several objects with \textit{z} $\mathrm{>}$ 7.  The quasar with the highest redshift at $z=7.5$, which corresponds to the age of the Universe at 650 million years, has an absolute luminosity $1.4\cdot 10^{47} $ erg/sec, while the estimate of the mass from the velocity of gas in the quasar gives the value of $1,6\pm 0,4\cdot 10^{9} M_{\mathrm{\odot }} $ [5]. These observational data raise the question of the mechanism of formation and rapid growth of such objects in the early Universe.

The results of numerical m [6] impose a number of restrictions on the formation parameters of supermassive black holes. Thus, for example, it has been shown that light nuclei of black holes with mass $M\mathrm{\leqslant }10^{3} M_{\mathrm{\odot }} $ cannot grow to masses of the order of $10^{8} M_{\mathrm{\odot }} $ per $z=6$ even during supercritical accretion. The formation of supermassive black holes with masses $10^{8} \div 10^{9} M_{\mathrm{\odot }} $ requires heavier nuclei $M\mathrm{\sim }10^{4} \div 10^{6} M_{\mathrm{\odot }} $ and gas-rich galaxies containing quasars. However, at present there are no sufficiently convincing models for the appearance of such heavy nuclei in the early Universe. In addition, it has been found that the spatial density of luminous quasars decreases rapidly with increasing redshift, with this trend increasing beyond \textit{z} = 5--6  [7].

Interest in the mechanisms of formation of supermassive Black Holes with masses $10^{8} -10^{9} M_{\mathrm{\odot }} $, taking into account the fact of their dominant presence in the composition of quasars, is caused, in particular, by the fact that such Black Holes are formed in the composition of quasars at fairly early stages of the evolution of the Universe, before the formation of stars. This circumstance, in particular, opens up the possibility of the formation of supermassive black holes under conditions where scalar fields and baryonic dark matter can significantly influence this process. In this regard, we note the works [8], [9] and [10], in which the possibility of the existence of scalar halos and scalar hairs in the vicinity of supermassive black holes is considered.

In [11], based on the developed theory of instability of a one-component system of scalarly charged degenerate fermions with a singlet Higgs scalar interaction in the so-called hard WKB approximation, the assumption of the instability of short-wavelength perturbations was confirmed. These preliminary studies showed the need for a comprehensive and deeper study of statistical systems of scalarly charged particles.

Further, in [12], two of the simplest models of the interaction of fermions with an asymmetric scalar doublet were proposed: in the first model, such an interaction is carried out by two types of fermions of different sorts, one of which is the source of the canonical scalar field, and the second is the phantom one (model$\mathrm{{\mathfrak M}}_{1} $); in the second model, there is one kind of fermions with a paired charge - canonical and phantom (model$\mathrm{{\mathfrak M}}_{2} $). A qualitative analysis of the dynamic systems of the model $\mathrm{{\mathfrak M}}_{1} $ was also carried out there.\textbf{}

Note, firstly, that in the case of a cosmological model based on the classical vacuum scalar singlet, the Hubble parameter is a nonincreasing function of time ($\dot{H}\mathrm{\leqslant }0$). In this case, inflationary solutions $H={\rm Const}$are possible only for constant values of the potential of the scalar field, corresponding to the points of stable equilibrium of the dynamical system. However, it turns out that such solutions are stable only with respect to infinitesimal perturbations. Therefore, despite the well-known energy and quantum-field problems with the singlet phantom field, we include the phantom field in the model studied below. At the same time, the transition to a singlet model based on a purely classical or purely phantom field is controlled by the parameters of the model, which, for theoretical completeness, are kept arbitrary by us, which allows us to identify the features of the influence of the model components on its properties.

Secondly, we note that we are considering a model without radiation, in which matter is represented by a cold degenerate system of scalarly charged fermions and scalar Higgs fields corresponding to these charges. Such a cold system of fermions can be considered as a model of dark matter in the early stages of the evolution of the Universe. In this paper, we study the gravitational stability of a model $\mathrm{{\mathfrak M}}_{1} $of a two-component statistical system with an asymmetric scalar interaction of fermions, removing the condition of the rigid WKB approximation [11].

\section{Mathematical model of two-component system of degenerate scalarly charged fermions with asymmetric Higgs interaction }

It was shown in [12] how, based on the Lagrangian formalism, from the microscopic equations of motion of scalarly charged particles, one can obtain a macroscopic model of a statistical system of scalarly charged particles described by macroscopic flows. In this article, we use the results obtained in [12].

Below, we will consider a cosmological model based on a two-component degenerate statistical system of scalarly charged fermions and an asymmetric scalar Higgs doublet consisting of a canonical scalar field $\Phi$ and a phantom field $\varphiup$ [14]. According to [12], the dynamical masses $m_{z} $ and $m_{\zeta } $fermions $z$ and  $\zeta $with charges $e_{z} $and $e_{\zeta } $with respect to the canonical field and the phantom field are described by the formulas:
\begin{equation} \label{GrindEQ__1_}
m_{z} =e_{z} \Phi ,\quad m_{\zeta } =e_{\zeta } \varphi .
\end{equation}
The Lagrange function $L_{s} $of non-interacting scalar Higgs fields of an asymmetric scalar doublet is
\begin{equation} \label{GrindEQ__2_}
L_{s} \equiv L_{\Phi } +L_{\varphi } =\frac{1}{16\pi } (g^{ik} \Phi _{,i} \Phi _{,k} -2V(\Phi ))+\frac{1}{16\pi } (-g^{ik} \varphi _{,i} \varphi _{,k} -2\mathrm{{\mathcal V}}(\varphi )),
\end{equation}
where  $V(\Phi )=-\frac{\alpha }{4} \left(\Phi ^{2} -\frac{m^{2} }{\alpha } \right)^{2} ,\quad \mathrm{{\mathcal V}}(\Phi )=-\frac{\beta }{4} \left(\varphi ^{2} -\frac{\mathrm{{\mathfrak m}}^{2} }{\beta } \right)^{2} $ \eqref{GrindEQ__3_} are the potential energies of the corresponding scalar fields, $\alphaup$  and $\betaup$ are their self-action constants, $m,\mathrm{{\mathfrak m}}$are their masses of quanta.

 The momentum energy tensor relative to the Lagrange function \eqref{GrindEQ__2_}, is as follows
\begin{equation} \label{GrindEQ__4_}
T_{(s)k}^{i} \equiv T_{(\Phi )\; k}^{i} +T_{(\varphi )\; k}^{i} =\frac{1}{16\pi } (2\Phi ^{,i} \Phi _{,k} \delta _{k}^{i} \Phi _{,j} \Phi ^{,j} +2V(\Phi )\delta _{k}^{i} )+\frac{1}{16\pi } (-2\varphi ^{,i} \varphi _{,k} \delta _{k}^{i} \varphi _{,j} \varphi ^{,j} +2\mathrm{{\mathcal V}}(\varphi )\delta _{k}^{i} ).
\end{equation}

Further, the energy-momentum tensor of an equilibrium statistical system is equal to:
\begin{equation} \label{GrindEQ__5_}
T_{(p)k}^{i} =(\varepsilon _{p} +p_{p} )u^{i} u_{k} -\delta _{k}^{i} p,
\end{equation}
where \textit{u${}^{i}$} is the vector of the macroscopic velocity of the statistical system.

Einstein's equations for the system ``scalar fields + particles'' have the form:
\begin{equation} \label{GrindEQ__6_}
R_{k}^{i} -\frac{1}{2} \delta _{k}^{i} R=8\pi T_{k}^{i} +\delta _{k}^{i} \Lambda _{0} ,
\end{equation}
\[T_{k}^{i} =T_{(s)k}^{i} +T_{(p)k}^{i} ,\]
where $\Lambda$${}_{0}$ is the seed value of the cosmological constant, associated with its observed value $\Lambda$, obtained by removing the constant terms in the potential energy, by the relation:
\begin{equation} \label{GrindEQ__7_}
\Lambda =\Lambda _{0} -\frac{1}{4} \sum _{r} \frac{m_{r}^{4} }{\alpha _{r} } .
\end{equation}

Strict macroscopic consequences of the kinetic theory, are transfer equations, including the conservation law of some vector current corresponding to the macroscopic conservation law in the reactions of some fundamental charge with multicharges $q_{(a)}^{r} $of particles
\begin{equation} \label{GrindEQ__8_}
\nabla _{i} \sum _{a} q_{(a)}^{r} n_{(a)}^{i} =0,
\end{equation}
and conservation laws of the momentum energy of the statistical system:
\begin{equation} \label{GrindEQ__9_}
\nabla _{k} T_{p}^{ik} -\sum _{r} \sigma ^{r} \nabla ^{i} \Phi _{r} =0,
\end{equation}
where $\sigmaup$\textit{${}^{r}$} is the density of scalar charges relative to the field $\Phi _{r} $ [12].

The well-known identity follows from the velocity vector normalization relation:
\begin{equation} \label{GrindEQ__10_}
u_{\; ,i}^{k} u_{k} \equiv 0,
\end{equation}
which allows reducing the momentum energy conservation law \eqref{GrindEQ__9_} to the equation of ideal hydrodynamics
\begin{equation} \label{GrindEQ__11_}
(\varepsilon _{p} +p_{p} )u_{\; ,k}^{i} u^{k} =(g^{ik} -u^{i} u^{k} )(p_{p,k} +\sum _{r} \sigma ^{r} \Phi _{r,k} ),
\end{equation}
\begin{equation} \label{GrindEQ__12_}
\nabla _{k} [(\varepsilon _{p} +p_{p} )u^{k} ]=u^{k} (p_{p,k} +\sum _{r} \sigma ^{r} \Phi _{r,k} ),
\end{equation}
and the conservation law of the fundamental charge \textit{Q${}^{r}$} \eqref{GrindEQ__8_} to the equation
\begin{equation} \label{GrindEQ__13_}
\nabla _{k} \rho ^{r} u^{k} =0,\quad (r=\overline{1,N}),
\end{equation}
where $\rho ^{r} \equiv \sum _{a} q_{(a)}^{r} n_{(a)} $ r\eqref{GrindEQ__14_} is the kinematic density of the scalar charge of the statistical system relative to the scalar field $\Phi$${}_{r}$.

Macroscopic scalars of the two-component statistical system of degenerate fermions takes the from:
\begin{equation} \label{GrindEQ__15_}
\varepsilon _{p} =\frac{e_{z}^{4} \Phi ^{4} }{8\pi ^{2} } F_{2} (\psi _{z} )+\frac{e_{\zeta }^{4} \varphi ^{4} }{8\pi ^{2} } F_{2} (\psi _{\zeta } ),
\end{equation}
\begin{equation} \label{GrindEQ__16_}
p_{p} =\frac{e_{z}^{4} \Phi ^{4} }{24\pi ^{2} } (F_{2} (\psi _{z} )-4F_{1} (\psi _{z} ))+\frac{e_{\zeta }^{4} \varphi ^{4} }{24\pi ^{2} } (F_{2} (\psi _{\zeta } )-4F_{1} (\psi _{\zeta } )),
\end{equation}
\begin{equation} \label{GrindEQ__17_}
\sigma ^{z} =\frac{e_{z}^{4} \Phi ^{3} }{2\pi ^{2} } F_{1} (\psi _{z} ),\quad \sigma ^{\zeta } =\frac{e_{\zeta }^{4} \varphi ^{3} }{2\pi ^{2} } F_{1} (\psi _{\zeta } ),
\end{equation}
where $\sigma ^{z} $ and $\sigma ^{\zeta } $ are densities of scalar charges $e_{z} $ and $e_{\zeta } $, respectively:
\begin{equation} \label{GrindEQ__18_}
\psi _{z} =\frac{\pi _{z} }{|e_{z} \Phi |} ,\quad \psi _{\zeta } =\frac{\pi _{\zeta } }{|e_{\zeta } \varphi |} .
\end{equation}

Functions $F_{1} (\psi )$ and $F_{2} (\psi )$ are introduced for simplicity:
\begin{equation} \label{GrindEQ__19_}
F_{1} (\psi )=\psi \sqrt{1+\psi ^{2} } -\ln (\psi +\sqrt{1+\psi ^{2} } ),
\end{equation}
\begin{equation} \label{GrindEQ__20_}
F_{2} (\psi )=\psi \sqrt{1+\psi ^{2} } (1+2\psi ^{2} )-\ln (\psi +\sqrt{1+\psi ^{2} } ).
\end{equation}
Functions $F_{1} (x)$ and $F_{2} (x)$, are odd functions
\begin{equation} \label{GrindEQ__21_}
F_{1} (-x)=-F_{1} (x),\quad F_{2} (-x)=-F_{2} (x),
\end{equation}
and have the following asymptotics:
\[\left. F_{1} (x)\right|_{x\to 0} \mathrm{\simeq }\frac{2}{3} x^{3} ,\; \; \left. F_{2} (x)\right|_{x\to 0} \mathrm{\simeq }\frac{8}{3} x^{3} ,\]
\begin{equation} \label{GrindEQ__22_}
\left. (F_{2} (x)-4F_{1} (x))\right|_{x\to 0} \mathrm{\simeq }\frac{8}{5} x^{5} ,
\end{equation}
\begin{equation} \label{GrindEQ__23_}
\left. F_{1} (x)\right|_{x\to \pm \infty } \mathrm{\simeq }x|x|,\quad \left. F_{2} (x)\right|_{x\to \pm \infty } \mathrm{\simeq }2x^{3} |x|.
\end{equation}
Next, we write derivative functions $F_{1} (x)$ and $F_{2} (x)$ useful for further relations:
\begin{equation} \label{GrindEQ__24_}
F'_{1} (x)=\frac{2x^{2} }{\sqrt{1+x^{2} } } ,\quad F'_{2} (x)=8x^{2} \sqrt{1+x^{2} } ,\quad \frac{d}{dx} x^{3} \sqrt{1+x^{2} } =3x^{2} \sqrt{1+x^{2} } +\frac{x^{4} }{\sqrt{1+x^{2} } } .
\end{equation}

Finally, equations of scalar fields for the two-component system take the form
\begin{equation} \label{GrindEQ__25_}
\mathrm{\square }\Phi +m^{2} \Phi -\alpha \Phi ^{3} =-\frac{4}{\pi ^{2} } e_{z}^{4} \Phi ^{4} F_{1} (\psi _{z} ),
\end{equation}
\begin{equation} \label{GrindEQ__26_}
-\mathrm{\square }\varphi +\mathrm{{\mathfrak m}}^{2} \varphi -\beta \varphi ^{3} =-\frac{4}{\pi ^{2} } e_{\zeta }^{4} \varphi ^{4} F_{1} (\psi _{\zeta } ).
\end{equation}

\section{Linear plane perturbations of cosmological model}

\subsection{Unperturbed isotropic homogeneous ground state}

Let us consider the spatial plane Friedman
\begin{equation} \label{GrindEQ__27_}
ds_{0}^{2} =a^{2} (\eta )(d\eta ^{2} -dx^{2} -dy^{2} -dz^{2} )\equiv dt^{2} -a^{2} (t)(dx^{2} +dy^{2} +dz^{2} ),
\end{equation}
where the cosmological time \textit{t }is associated with the time variable $\etaup$ in the following way:
\begin{equation} \label{GrindEQ__28_}
t=\int  a(\eta )d\eta .
\end{equation}
As a background solution, let us consider the uniform isotropic distribution of the matter, in which all thermodynamic functions and scalar fields are dependent on the time only:
\begin{equation} \label{GrindEQ__29_}
S0:\quad \Phi =\Phi (t),\; \varphi =\varphi (t),\; \pi _{z} =\pi _{z} (t),\; \pi _{\zeta } =\pi _{\zeta } (t),\; u^{i} =u^{i} (t).
\end{equation}
It is easy to verify that
\begin{equation} \label{GrindEQ__30_}
u^{i} =\delta _{4}^{i}
\end{equation}
turns equations \eqref{GrindEQ__11_} into identities, whereas Eqns \eqref{GrindEQ__12_}, \eqref{GrindEQ__13_} are reduced to three constitutive equations:
\begin{equation} \label{GrindEQ__31_}
\dot{\varepsilon }_{p} +3\frac{\dot{a}}{a} (\varepsilon _{p} +p_{p} )=\sigma ^{z} \dot{\Phi }+\sigma ^{\zeta } \dot{\varphi },
\end{equation}
\begin{equation} \label{GrindEQ__32_}
\dot{n}_{z} +3\frac{\dot{a}}{a} n_{z} =0,
\end{equation}
\begin{equation} \label{GrindEQ__33_}
\dot{n}_{\zeta } +3\frac{\dot{a}}{a} n_{\zeta } =0,
\end{equation}
where $\dot{\phi }\equiv d\phi /dt$. It was shown in [12] that the system of equations \eqref{GrindEQ__31_} -- \eqref{GrindEQ__33_} has simple exact solutions:
\begin{equation} \label{GrindEQ__34_}
a\pi _{z} ={\rm const,}\quad a\pi _{\zeta } ={\rm const}.
\end{equation}
Taking into account \eqref{GrindEQ__34_} and \eqref{GrindEQ__1_}, we write the dimensionless functions and \eqref{GrindEQ__18_} in an explicit form:
\begin{equation} \label{GrindEQ__35_}
\psi _{z} =\frac{\pi _{z}^{0} }{a|e_{z} \Phi |} {\rm e}^{-\xi } ,\quad \psi _{\zeta } =\frac{\pi _{\zeta }^{0} }{a|e_{\zeta } \varphi |} {\rm e}^{-\xi } \quad (\pi _{(a)}^{0} =\pi _{(a)} (0)),
\end{equation}
where we have moved to a new dimensionless variable $\xi (t)$\underbar{}
\begin{equation} \label{GrindEQ__36_}
\xi =\ln a,
\end{equation}
assuming here and in the future
\begin{equation} \label{GrindEQ__37_}
\xi (0)=0.
\end{equation}

The momentum energy tensor of the scalar field in the non-perturbates state, also takes the form of the momentum energy tensor of the true isotropic fluid:
\begin{equation} \label{GrindEQ__38_}
T_{s}^{ik} =(\varepsilon _{s} +p_{s} )u^{i} u^{k} -p_{s} g^{ik} ,
\end{equation}
at
\begin{equation} \label{GrindEQ__39_}
\varepsilon _{s} =\frac{1}{8\pi } \left(\frac{1}{2} \frac{\dot{\Phi }^{2} -\dot{\varphi }^{2} }{a^{2} } +V(\Phi )+\mathrm{{\mathcal V}}(\varphi )\right),
\end{equation}
\begin{equation} \label{GrindEQ__40_}
p_{s} =\frac{1}{8\pi } \left(\frac{1}{2} \frac{\dot{\Phi }^{2} -\dot{\varphi }^{2} }{a^{2} } -V(\Phi )-\mathrm{{\mathcal V}}(\varphi )\right),
\end{equation}
therefore
\begin{equation} \label{GrindEQ__41_}
\varepsilon _{s} +p_{s} =\frac{{\rm e}}{8\pi } \frac{\dot{\Phi }^{2} -\dot{\varphi }^{2} }{a^{2} } .
\end{equation}
In the Friedman--Robertson--Walker metric, Eqns \eqref{GrindEQ__25_}, \eqref{GrindEQ__26_} of non-perturbated scalar fields take the form
\begin{equation} \label{GrindEQ__42_}
\ddot{\Phi }+3\frac{\dot{a}}{a} \dot{\Phi }+m^{2} \Phi -\alpha \Phi ^{3} =-8\pi a^{2} \sigma ^{z} ,
\end{equation}
\begin{equation} \label{GrindEQ__43_}
\ddot{\varphi }+3\frac{\dot{a}}{a} \dot{\varphi }-\mathrm{{\mathfrak m}}^{2} \varphi +\beta \varphi ^{3} =8\pi a^{2} \sigma ^{\zeta } ,
\end{equation}
where densities $\sigma ^{z} $ and $\sigma ^{\zeta } $ of scalar charges are described by Eqns \eqref{GrindEQ__17_}, in which it is necessary to insert functions $\psi _{z} $ and $\psi _{\zeta } $ from Eq. \eqref{GrindEQ__35_}.

 Independent Einstein equations of zeroth-order approximation can be written as [13]:
\begin{equation} \label{GrindEQ__44_}
\dot{H}=-\frac{\dot{\Phi }^{2} }{2} +\frac{\dot{\varphi }^{2} }{2} -\frac{4}{3\pi } e_{z}^{4} \Phi ^{4} \psi _{z}^{3} \sqrt{1+\psi _{z}^{2} } -\frac{4}{3\pi } e_{\zeta }^{4} \varphi ^{4} \psi _{\zeta }^{3} \sqrt{1+\psi _{\zeta }^{2} } ,
\end{equation}
\begin{equation} \label{GrindEQ__45_}
3H^{2} -\Lambda -\frac{\dot{\Phi }^{2} }{2} +\frac{\dot{\varphi }^{2} }{2} -\frac{m^{2} \Phi ^{2} }{2} +\frac{\alpha \Phi ^{4} }{4} -\frac{\mathrm{{\mathfrak m}}^{2} \varphi ^{2} }{2} +\frac{\beta \varphi ^{4} }{4} -\frac{e_{z}^{4} \Phi ^{4} }{\pi } F_{2} (\psi _{z} )-\frac{e_{\zeta }^{4} \varphi ^{4} }{\pi } F_{2} (\psi _{\zeta } )=0,
\end{equation}
Where \textit{H}($\etaup$) is the Hubble constant
\begin{equation} \label{GrindEQ__46_}
H=\frac{\dot{a}}{a} \equiv \dot{\xi }.
\end{equation}

Thus, Eqns \eqref{GrindEQ__42_}, \eqref{GrindEQ__43_} and \eqref{GrindEQ__46_} compose a full system of ordinary differential equations relative to $\xi (t)$, $H(t)$, $\Phi (t)$ functions, which describe the non-perturbated cosmological model. As shown in [12], for example, Eq. \eqref{GrindEQ__45_} is the first integral of this system, which helps to determine the initial value of the Hubble constant when solving the Cauchy problem.

Note that this system of equations is more suitable to investigate in the time scale of physical time \textit{t} (Eq. (28)), as relative to this variable, the fully autonomous system of equations for the cosmological model, takes a simpler form [12]
\begin{equation} \label{GrindEQ__47_}
\dot{\xi }=H,\; \; \dot{\Phi }=Z,\; \; \dot{\varphi }=z,
\end{equation}
\begin{equation} \label{GrindEQ__48_}
\dot{H}=-\frac{Z^{2} }{2} +\frac{z^{2} }{2} -\frac{4}{3\pi } e_{z}^{4} \Phi ^{4} \psi _{z}^{3} \sqrt{1+\psi _{z}^{2} } -\frac{4}{3\pi } e_{\zeta }^{4} \varphi ^{4} \psi _{\zeta }^{3} \sqrt{1+\psi _{\zeta }^{2} } ,
\end{equation}
\begin{equation} \label{GrindEQ__49_}
\dot{Z}=-3HZ-m^{2} \Phi +\Phi ^{3} \left(\alpha -\frac{4e_{z}^{4} }{\pi } F_{1} (\psi _{z} )\right),
\end{equation}
\begin{equation} \label{GrindEQ__50_}
\dot{z}=-3Hz+\mathrm{{\mathfrak m}}^{2} \varphi -\varphi ^{3} \left(\beta -\frac{4e_{\zeta }^{4} }{\pi } F_{1} (\psi _{\zeta } )\right).
\end{equation}
The relation for the first integral in Eq. \eqref{GrindEQ__45_} can be rewritten as
\begin{equation} \label{GrindEQ__51_}
3H^{2} -\Lambda -\frac{Z^{2} }{2} +\frac{z^{2} }{2} -\frac{m^{2} \Phi ^{2} }{2} +\frac{\alpha \Phi ^{4} }{4} -\frac{\mathrm{{\mathfrak m}}^{2} \varphi ^{2} }{2} +\frac{\beta \varphi ^{4} }{4} -\frac{e_{z}^{4} \Phi ^{4} }{\pi } F_{2} (\psi _{z} )-\frac{e_{\zeta }^{4} \varphi ^{4} }{\pi } F_{2} (\psi _{\zeta } )=0.
\end{equation}

\section{First-order equations of perturbation}

Below we also use the time variable $\etaup$ for the adequacy to the standard theory of Lifshitz perturbations (e.g., [13]). Using $f'$, we denote the time variable derivative. It is necessary to allow for simple rules of the differentiation
\begin{equation} \label{GrindEQ__52_}
\dot{\phi }=\frac{1}{a} \phi ',\; \; \ddot{\phi }=\frac{1}{a^{2} } \phi ''-\frac{a'}{a^{3} } \phi ',\Rightarrow \phi '=a\dot{\phi },\; \; \phi ''=a^{2} \ddot{\phi }+a\dot{a}\dot{\phi }
\end{equation}
and relation $\xi '=aH$.

\subsection{Longitudinal perturbations}

The metric with gravitation perturbations can be written as follows (e.g., [13]):
\begin{equation} \label{GrindEQ__53_}
ds^{2} =ds_{0}^{2} -a^{2} (\eta )h_{\alpha \beta } dx^{\alpha } dx^{\beta } .
\end{equation}
Note that conformal multiplier $-a^{2} (\eta )$ before covariant perturbation amplitudes disappears for the mixed components of perturbations$h_{\beta }^{\alpha } $. Covariant perturbations of the metric can be found from
\begin{equation} \label{GrindEQ__54_}
\delta g_{\alpha \beta } =-a^{2} (t)h_{\alpha \beta } .
\end{equation}
Next,
\begin{equation} \label{GrindEQ__55_}
h_{\beta }^{\alpha } =h_{\gamma \beta } g_{0}^{\alpha \gamma } \equiv -\frac{1}{a^{2} } h_{\alpha \beta } ,
\end{equation}
\begin{equation} \label{GrindEQ__56_}
h\equiv h_{\alpha }^{\alpha } \equiv g_{0}^{\alpha \beta } h_{\alpha \beta } =-\frac{1}{a^{2} } (h_{11} +h_{22} +h_{33} ).
\end{equation} 

Further, we consider only longitudinal perturbations of the metric implying the problem of the gravitational stability of plane perturbations and directing the wave vector along the \textit{Oz} axis. In this, coordinate system,
\[h_{11} =h_{22} =\frac{1}{3} [\lambda (t)+\frac{1}{3} \mu (t)]{\rm e}^{inz} ,\]
\[h=\mu (t){\rm e}^{inz} ,\; \; h_{12} =h_{13} =h_{23} =0,\]
\begin{equation} \label{GrindEQ__57_}
h_{33} =\frac{1}{3} [-2\lambda (t)+\mu (t)]{\rm e}^{inz} .
\end{equation}

One can see that the matter in our model is completely determined by four scalar functions, \textit{viz}. $\Phi (z,\eta )$, $\varphi (z,\eta )$, $\pi _{z} (z,\eta )$, and $\pi _{\zeta } (z,\eta )$ and the velocity vector $u^{i} (z,\eta )$. According to [11], expand these functions by the smallness of perturbations relative to the respective functions on the background of the Friedman metric \eqref{GrindEQ__53_}\footnote{ $\ For\ simplicity,\ accept\ $  $S_{0} =S(t)$  $\ for\ non-perturbated\ values\ $  $S_{0} (t)$  $\ $ }:
\[\Phi (z,\eta )=\Phi (\eta )+\delta \Phi (\eta ){\rm e}^{inz} ,\; \; \varphi (z,\eta )=\varphi (\eta )+\delta \varphi (\eta ){\rm e}^{inz} ,\]
\begin{equation} \label{GrindEQ__58_}
\pi _{z} (z,t)=\pi _{z} (\eta )(1+\delta _{z} (\eta ){\rm e}^{inz} ),\; \; \pi _{\zeta } (z,t)=\pi _{\zeta } (\eta )(1+\delta _{\zeta } (\eta ){\rm e}^{inz} ),
\end{equation}
\[\sigma ^{z} (z,\eta )=\sigma ^{z} (\eta )+\delta \sigma ^{z} (\eta ){\rm e}^{inz} ,\; \; \sigma ^{\zeta } (z,\eta )=\sigma ^{\zeta } (\eta )+\delta \sigma ^{\zeta } (\eta ){\rm e}^{inz} ,\]
\[u^{i} =\frac{1}{a} \delta _{4}^{i} +\delta _{3}^{i} \mathit{v}(\eta ){\rm e}^{inz} ,\]
where $\delta \Phi (\eta )$, $\delta \varphi (\eta )$, $\delta _{z} (\eta ),\delta _{\zeta } (\eta )$, $s_{z} (\eta ),s_{\zeta } (\eta )$, and $\mathit{v}(\eta )$ are first-order smallness functions compared to their non-perturbated values.

\subsection{Equations of perturbation for scalar fields }

The Taylor expansion of field equations \eqref{GrindEQ__25_} and \eqref{GrindEQ__26_} by the smallness of perturbations, yields equations of perturbations of the first-order scalar fields $\delta \Phi $, $\delta \varphi $:
\begin{equation} \label{GrindEQ__59_}
\delta \Phi ''+2\frac{a'}{a} \delta \Phi '+[n^{2} +a^{2} (m^{2} -3\alpha \Phi ^{2} )]\delta \Phi +\frac{1}{2} \Phi '\mu '=-8\pi a^{2} \delta \sigma ^{z} ,
\end{equation}
\begin{equation} \label{GrindEQ__60_}
\delta \varphi ''+2\frac{a'}{a} \delta \varphi '+[n^{2} -a^{2} (\mathrm{{\mathfrak m}}^{2} +3\beta \varphi ^{2} )]\delta \varphi +\frac{1}{2} \varphi '\mu '=8\pi a^{2} \delta \sigma ^{\zeta } .
\end{equation}
These equations differ from similar equations of vacuum scalar field perturbations by only the term with the scalar field source on the right-hand side and signs in kinetic terms for the phantom field [14].

\noindent
{\it 3.3. Equations for gravitational perturbations}

After the Taylor expansion of the Einstein equation \eqref{GrindEQ__6_} by perturbation orders (see [11]), we get the following independent equations for the first-order gravitational perturbations µ and $\lambdaup$:
\begin{equation} \label{GrindEQ__61_}
\mathit{v}=\frac{in}{8\pi a^{3} (\varepsilon +p)_{p} } (\delta \Phi \Phi '-\delta \varphi \varphi '+\frac{1}{3} (\lambda '+\mu ')),
\end{equation}
\begin{equation} \label{GrindEQ__62_}
8\pi a^{2} \delta \varepsilon _{p} =\frac{a'}{a} \mu '-\Phi '\delta \Phi '+\varphi '\delta \varphi '+\frac{n^{2} }{3} (\lambda +\mu )-a^{2} (m^{2} -\alpha \Phi ^{2} )\Phi \delta \Phi -a^{2} (\mathrm{{\mathfrak m}}^{2} -\beta \varphi ^{2} )\varphi \delta \varphi ,
\end{equation}
\begin{equation} \label{GrindEQ__63_}
\lambda ''+2\frac{a'}{a} \lambda '-\frac{1}{3} n^{2} (\lambda +\mu )=0,
\end{equation}
\begin{equation} \label{GrindEQ__64_}
\mu ''+2\frac{a'}{a} \mu '+\frac{1}{3} n^{2} (\lambda +\mu )+3\delta \Phi '\Phi '-3\delta \varphi '\varphi '-3a^{2} [\Phi \delta \Phi (m^{2} -\alpha \Phi ^{2} )+\varphi \delta \varphi (m^{2} -\beta \varphi ^{2} )-8\pi \delta p_{p} ]=0.
\end{equation}
We can show (e.g., [11]), that incremental and algebraic consequences of Eqns \eqref{GrindEQ__61_}--\eqref{GrindEQ__64_} are equations \eqref{GrindEQ__59_}, \eqref{GrindEQ__60_} for perturbations of the scalar field and the motion of the first approximation of the degenerated matter. The latter are obviously excessive, because perturbations of the velocity and energy density of matter are directly determined by Eqns \eqref{GrindEQ__61_} and \eqref{GrindEQ__62_}. Let us select independent equations \eqref{GrindEQ__59_}, \eqref{GrindEQ__60_}, \eqref{GrindEQ__63_} and \eqref{GrindEQ__64_} from the remained equations in order to approximate the mathematical model to the standard Lifshitz theory. Equation \eqref{GrindEQ__63_} coincides with the respective equation of the Lifshitz theory, whereas Eq. \eqref{GrindEQ__64_} differs from the respective equation from [14] by only the material term $\deltaup$\textit{p}${}_{p}$ and additions of the similar term for the phantom field.

Note that in the absence of the fermion system $\varepsilon _{p} =\delta \varepsilon _{p} =p_{p} =\delta p_{p} =0$, the expression on the right-hand side of Eq. \eqref{GrindEQ__61_} becomes zero and yields the equation of the field and metric perturbations similar to Eq. \eqref{GrindEQ__62_}, which instead of the perturbation determination of the energy density of the fermion component, becomes the equation of perturbations. According to [14], not all equations of vacuum scalar field and metric perturbations, are independent. In [15], WKB solutions are found for the respective equations of classical scalar field perturbations. Since the system of perturbation equations of vacuum scalar fields differ from the system of perturbation equations of fields with sources, we assume the obligatory presence of the fermion component and the following conditions:
\begin{equation} \label{GrindEQ__65_}
\varepsilon _{p} \rlap{$/$}\equiv 0;\quad \delta \varepsilon _{p} \rlap{$/$}\equiv 0.
\end{equation}

Introducing a new variable for gravitation perturbations
\begin{equation} \label{GrindEQ__66_}
\nu =\lambda +\mu
\end{equation}
and summing both members of equations \eqref{GrindEQ__63_} and \eqref{GrindEQ__64_}, we get the new system of equations for variables $\lambda $ and $\nu $:
\begin{equation} \label{GrindEQ__67_}
\lambda ''+2\frac{a'}{a} \lambda '-\frac{1}{3} n^{2} \nu =0,
\end{equation}
\begin{equation} \label{GrindEQ__68_}
\nu ''+2\frac{a'}{a} \nu '+\frac{1}{3} n^{2} \nu +3\delta \Phi '\Phi '-3\delta \varphi '\varphi '-3a^{2} [\Phi \delta \Phi (m^{2} -\alpha \Phi ^{2} )+\varphi \delta \varphi (\mathrm{{\mathfrak m}}^{2} -\beta \varphi ^{2} )-8\pi \delta p_{p} ]=0.
\end{equation}

To close the system of equations, it is necessary to find relations between macroscopic scalars on the one hand, and perturbations of scalar and gravitation fields, on the other hand.

\noindent
{\it 3.4. Perturbations of fermion component }

Let us find the indicated above relation between macroscopic scalars of the degenerated Fermi matter with perturbations of scalar and gravitation fields. With regard to Eq. \eqref{GrindEQ__5_} true for the ideal fluid [11] and reporting perturbations of the Fermi momentum in terms of
\begin{equation} \label{GrindEQ__69_}
\delta \pi _{z} \equiv \pi _{z} \delta _{z} (\eta ){\rm e}^{inz} ,\quad \delta \pi _{\zeta } \equiv \pi _{\zeta } \delta _{\zeta } (\eta ){\rm e}^{inz} ,
\end{equation}
write perturbations of macroscopic scalars in the first approximation:
\[\delta n_{z} =3n_{z} (t)\delta _{z} (t){\rm e}^{inz} ,\quad \delta \psi _{z} =\psi _{z} (\eta )\gamma _{z} (\eta ){\rm e}^{inz} ,\]
\begin{equation} \label{GrindEQ__70_}
\delta n_{\zeta } =3n_{\zeta } (t)\delta _{\zeta } (t){\rm e}^{inz} ,\quad \delta \psi _{\zeta } =\psi _{\zeta } (\eta )\gamma _{\zeta } (\eta ){\rm e}^{inz} ,
\end{equation}
where, according to Eq. \eqref{GrindEQ__18_},
\begin{equation} \label{GrindEQ__71_}
\psi _{z} (\eta )=\frac{\pi _{z} (\eta )}{|e_{z} \Phi (\eta )|} ,\; \; \psi _{\zeta } (\eta )=\frac{\pi _{\zeta } (\eta )}{|e_{\zeta } \varphi (\eta )|} ,\; \; n_{z} (t)=\frac{\pi _{z}^{3} (\eta )}{\pi ^{2} } ,\; \; n_{\zeta } (t)=\frac{\pi _{\zeta }^{3} (\eta )}{\pi ^{2} }
\end{equation}
and perturbations of the reduced Fermi momentum $p_{f} /m$ are given for each fermion species:
\begin{equation} \label{GrindEQ__72_}
\gamma _{z} (\eta )=\delta _{z} (\eta )-\frac{\delta \Phi (\eta )}{\Phi (\eta )} ,\quad \gamma _{\zeta } (\eta )=\delta _{\zeta } (\eta )-\frac{\delta \varphi (\eta )}{\varphi (\eta )} .
\end{equation}

Additionally, based on the charge conservation law for $e_{z} $ and $e_{\zeta } $ \eqref{GrindEQ__11_} and Eq. \eqref{GrindEQ__70_}, we can obtain conservation laws of the number of each fermion species in the first-order theory of perturbations:
\[\frac{1}{2} \mu '+3\delta '_{z} +in\mathit{v}=0,\quad \quad \frac{1}{2} \mu '+3\delta '_{\zeta } +in\mathit{v}=0,\]
therefore,
\begin{equation} \label{GrindEQ__73_}
\delta _{z} =\delta _{\zeta } \equiv \delta ,
\end{equation}
\begin{equation} \label{GrindEQ__74_}
\frac{1}{2} \mu '+3\delta '+in\mathit{v}=0,
\end{equation}
i.e., relative perturbations of the Fermi momentum of degenerated fermions coincide with each other. Equation \eqref{GrindEQ__74_} is an alternative (Eq. (61)) determination of the perturbation velocity \textit{v}. Note that both these equations are obtained as conservation laws for the transfer equations, i.e., Eq. \eqref{GrindEQ__74_} is derived from the charge conservation law, whereas Eq. \eqref{GrindEQ__61_} is derived from the momentum energy of the statistical system. As shown in [12], the fermion energy conservation law originates from the charge conservation law. Thus, the perturbation velocity \textit{v}($\etaup$) is determined algebraically through any of indicated relations. Equations \eqref{GrindEQ__72_} take the form
\begin{equation} \label{GrindEQ__75_}
\gamma _{z} (\eta )=\delta (\eta )-\frac{\delta \Phi (\eta )}{\Phi (\eta )} ,\quad \gamma _{\zeta } (\eta )=\delta (\eta )-\frac{\delta \varphi (\eta )}{\varphi (\eta )} .
\end{equation}

Further, taking into account differential identities \eqref{GrindEQ__24_}, we find expressions for perturbations of macroscopic fermion scalars (in accordance with the remark in Section 3.1 below, we omit the exponential factor ${\rm e}^{inz} $):
\begin{equation} \label{GrindEQ__76_}
\delta \sigma ^{z} =\frac{e_{z}^{4} \Phi ^{3} }{2\pi ^{2} } \left[\left(3F_{1} (\psi _{z} )-\frac{\psi _{z}^{3} }{\sqrt{1+\psi _{z}^{2} } } \right)\frac{\delta \Phi }{\Phi } +\frac{\psi _{z}^{3} }{\sqrt{1+\psi _{z}^{2} } } \delta \right],
\end{equation}
\begin{equation} \label{GrindEQ__77_}
\delta \sigma ^{\zeta } =\frac{e_{\zeta }^{4} \varphi ^{3} }{2\pi ^{2} } \left[\left(3F_{1} (\psi _{\zeta } )-\frac{\psi _{\zeta }^{3} }{\sqrt{1+\psi _{\zeta }^{2} } } \right)\frac{\delta \varphi }{\varphi } +\frac{\psi _{\zeta }^{3} }{\sqrt{1+\psi _{\zeta }^{2} } } \delta \right],
\end{equation}
\begin{equation} \label{GrindEQ__78_}
\delta \varepsilon _{p} =\frac{e_{z}^{4} \Phi ^{3} }{2\pi ^{2} } F_{1} (\psi _{z} )\delta \Phi +\frac{e_{\zeta }^{4} \varphi ^{3} }{2\pi ^{2} } F_{1} (\psi _{\zeta } )\delta \varphi +\frac{1}{\pi ^{2} } \left(e_{z}^{4} \Phi ^{4} \psi _{z}^{3} \sqrt{1+\psi _{z}^{2} } +e_{\zeta }^{4} \varphi ^{4} \psi _{\zeta }^{3} \sqrt{1+\psi _{\zeta }^{2} } \right)\delta ,
\end{equation}
\begin{equation} \label{GrindEQ__79_}
\delta p_{p} =\frac{e_{z}^{4} \Phi ^{3} }{2\pi ^{2} } F_{3} (\psi _{z} )\delta \Phi +\frac{e_{\zeta }^{4} \varphi ^{3} }{2\pi ^{2} } F_{3} (\psi _{\zeta } )\delta \varphi +\frac{1}{3\pi ^{2} } \left(e_{z}^{4} \Phi ^{4} \frac{\psi _{z}^{4} }{\sqrt{1+\psi _{z}^{2} } } +e_{\zeta }^{4} \varphi ^{4} \frac{\psi _{\zeta }^{4} }{\sqrt{1+\psi _{\zeta }^{2} } } \right)\delta ,
\end{equation}
where  $F_{3} (x)=\frac{x^{3} }{\sqrt{1+x^{2} } } +\ln (x+\sqrt{1+x^{2} } )\mathrm{\geqslant }0$.

As a result, all perturbations of macroscopic scalars for a degenerate plasma are completely determined by three perturbation functions: $\delta \Phi (\eta )$, $\delta \varphi (\eta )$ and $\delta (\eta )$.
Thus, substituting the expressions for macroscopic scalars \eqref{GrindEQ__76_} - \eqref{GrindEQ__79_} into the equations of scalar fields \eqref{GrindEQ__59_}, \eqref{GrindEQ__60_} and the Einstein equations \eqref{GrindEQ__62_}, \eqref{GrindEQ__67_} and \eqref{GrindEQ__68_}, we obtain the desired closed system of five ordinary differential equations second order with respect to five perturbation functions$\delta \Phi (\eta )$, $\delta \varphi (\eta )$, $\lambda (\eta )$, $\mu (\eta )$, $\delta (\eta )$. At the same time, the specificity of the variable $\delta (\eta )$lies in the fact that this variable enters the equations in a linear algebraic way, which allows, in principle, to exclude it through one of the equations and, thereby, reduce the system under study to a system of four differential equations with respect to four functions $\delta \Phi (\eta )$, $\delta \varphi (\eta )$, $\lambda (\eta )$, $\mu (\eta )$, $\delta (\eta )$. The easiest way to do this is by substituting the expression \eqref{GrindEQ__78_} into the Einstein equation \eqref{GrindEQ__62_}, the found expression for $\delta (\eta )$ is substituted into the scalars $s^{z} $ \eqref{GrindEQ__76_}, $s^{\zeta } $ \eqref{GrindEQ__77_}, $\delta p_{p} $ \eqref{GrindEQ__79_}, and then into the studied equations \eqref{GrindEQ__59_}, \eqref{GrindEQ__60_} and \eqref{GrindEQ__68_A20_}. However, we will not write out the equations obtained in this way because of their extremely cumbersome and of little use.

\subsection*{Funding}

This paper was supported by the Kazan Federal University Strategic Academic Leadership Program.

\end{document}